\documentclass[12pt,epsfig]{article}
\usepackage{graphicx}
\usepackage{amssymb}
\usepackage{mathrsfs}

\newcommand{\beq}[1] {\begin{equation}\label{#1} }
\newcommand{\eeq} {\end{equation} }
\newcommand{\bea}[1]{\begin{eqnarray}\label{#1} }
\newcommand{\eea}{\end{eqnarray}}

\begin{document}

\vspace*{-0.5cm}
\begin{flushright}
OSU-HEP-06-3
\end{flushright}
\vspace{0.5cm}

\begin{center}
{\Large {\bf A Model for Neutrino and Charged Lepton Masses in Extra Dimensions} }

\vspace*{1.5cm}
S. Nandi\footnote{Email address: s.nandi@okstate.edu}$^{,\dag}$ and
C. M. Rujoiu\footnote{Email address:
marius.rujoiu@okstate.edu}$^{,\dag,\ddag}$

\vspace*{0.5cm}
$^{\dag}${\it Department of Physics and Oklahoma Center for High Energy Physics,\\
Oklahoma State University, Stillwater, Oklahoma, 74078\\}

$^{\ddag}${\it Institute of Space Sciences, Bucharest-Magurele\\
Romania, 76900\\}

\end{center}

\begin{abstract}
We propose a model with one large submm size extra dimension in which the gravity and right-handed (RH) neutrino propagate, but the three Standard Model (SM) families are confined to fat branes of $TeV^{-1}$ size or smaller. The charged leptons and the light neutrinos receive mass from the five dimensional Yukawa couplings with the SM singlet neutrino via electroweak Higgs, while the KK excitations of the SM singlet neutrino gets large $TeV$ scale masses from the five dimensional Yukawa coupling with an electroweak singlet Higgs. The model gives non-hierarchical light neutrino masses, accommodate hierarchical charged lepton masses, and naturally explain why the light neutrino masses are so much smaller compared to the charged lepton masses. Large neutrino mixing is naturally expected in this scenario. The light neutrinos are Dirac particles in this model, hence neutrinoless double beta decay is not allowed. The model has also several interesting collider implications and can be tested at the LHC.
\end{abstract}

\section{Introduction}

In the past few years, atmospheric and solar neutrino experiments have demonstrated that the neutrinos do have very small masses, of the order of one hundredth of an electron volt [1-3]. These masses are a million or more times smaller than the corresponding charged lepton masses. While the charged lepton masses are highly hierarchical, with electron mass about half of a million electron volt to tau lepton mass about two thousand million electron volts, the neutrino masses are not much hierarchical. The square root of the neutrino mass square differences are about only one to five [1-3]. In addition, mixing among the neutrinos are large, compared to very small mixings among the quarks [1-3]. These observations have led to several unanswered questions. Why the neutrino masses are so small compared to the corresponding charged lepton masses? Why there is practically no hierarchy among the neutrino masses, while there are large hierarchies among the charged lepton masses? Also, unlike the quark sector why the mixing angles in the neutrino sector are large? Another related fundamental question is weather neutrinos are Majorana or Dirac particles.

The most popular idea proposed so far for understanding the tiny neutrino mass is the famous see-saw mechanism \cite{seesaw}. One postulates the existence of a very massive Standard Model (SM) singlet right handed neutrino with Majorana masses of order of $M \sim 10^{14}GeV$. The Yukawa coupling of the left-handed neutrino to this heavy right-handed neutrino then gives a Dirac mass of the order of the charged lepton masses, $m_l$. As a result, the left-handed neutrino obtains a tiny mass of the order of  $m^2_l/ M$. See-saw model naturally does not lead to lack of hierarchy among the light neutrino masses. In addition, in this model, the light neutrino masses are Majorana particles and will lead to neutrino-less double beta decay. Current limit on the neutrino-less double beta decay ($\beta\beta$) $m_{ee} \leq 0.3-1 eV$ \cite{betabeta}. What if no $\beta\beta$-decay is observed in the future precision experiments?

Recently, another interesting idea has been proposed to understand the smallness of the neutrino masses in the context of large extra dimensions (the ADD model) \cite{ADD}. Here, all the SM particles are confined to the four dimensional wall (three space and one time, the so called $D_3$ brane). There exists a SM singlet right-handed (RH) neutrino, which, like gravity, propagates into the sub-mm size extra dimension. The Yukawa couplings of the SM left-handed neutrino with this RH neutrino then gives a tiny Dirac mass for the LH neutrino. In this picture, the smallness of the neutrino mass arises from the smallness of the effective four dimensional Yukawa coupling due to large volume of the extra dimension in which the RH neutrino propagates. Several extensions of this idea to study the neutrino masses and mixings in detail have been pursued \cite{dics}.

In this work, we present a model which can explain why there are large hierarchy among the charged lepton masses, but practically no hierarchy among the light neutrino masses. The light neutrinos get a tiny mass as in the ADD mechanism; but their masses are related to the corresponding charged lepton masses. The neutrinos are naturally Dirac particles. When extended to accommodate two or three families, the model naturally provides large mixings among neutrinos. Our model unifies the ADD scenario \cite{nima} with the scenario of the Universal Extra Dimensions (UED) \cite{acd}. The SM particles of one family live in a fat brane \cite{rigolin},\cite{rius} of $TeV^{-1}$ size or smaller which is a tiny part of the submm size extra dimensions. The graviton as well as the SM singlet neutrinos propagate in the submm size extra dimensions. The three SM fermion families live in fat branes of three different sizes ($R_1$, $R_2$, $R_3$). Since all the charged leptons get their masses from the Yukawa couplings in five dimensions of sizes ($R_1$, $R_2$, $R_3$), their mass hierarchies are naturally obtained from the hierarchy of the sizes of the fat branes. Note that our model is different from the split fermions in extra dimensions model [12] where the hierarchy of the charged fermion masses are obtained from the different overlaps of the SM Higgs field wave function to the split fermion locations. Light neutrinos, on the other hand, as in ADD, get their masses from their Yukawa couplings with the SM singlet bulk neutrino propagating in the submm size extra dimensions. So, their masses are of the same order, and hence there is no hierarchy among the light neutrino masses. The neutrino masses are much smaller compared to the charged lepton masses because their Yukawa interactions get diluted by the volume of the submm size dimensional space compared to these of the charged leptons. Since the light neutrino masses are all of same order including their off-diagonal Yukawa coupling, large mixing among the neutrinos naturally arise.

Our work is presented as follows: In section 2, we present the model and the formalism and the calculation for the charged lepton and neutrino masses for one family. In section 3, we generalize this to more families, and discuss the resulting neutrino masses and mixings. The phenomenological implications of the model is described in section 4. Section 5 contains our conclusions.

\section{Model and the formalism}

We consider one extra dimension of size $r$ (which, as in ADD model, is of submm size). This extra dimension is taken to be $S^1$ and is denoted by $y$. The gravity propagates all the way in this extra dimension. The SM singlet neutrino can propagate in $y$ only in the space given by $S^1/Z_2$ orbifold. All the SM particles of the first family can propagate only up to a distance $R$ within the $5^{th}$ dimension of size $r$. Thus, these SM particles live in a fat brane of thickness $R$, and the space in which they propagate is $S^1/Z_2$ orbifold of size $R$. We denote the left-handed lepton doublet $(\nu_e,e)_L$ by $l(x,y)$, RH lepton singlet singlet by $l_R(x,y)$, Higgs doublet by $H(x,y)$, and the SM singlet neutrino by $N(x,y)$. Note that $l$, $l_R$ and $H$ are confined to the fat brane of size $R$ (of $TeV^{-1}$ size), while $N$ and the gravity propagates in size $r$. We can expand the five dimensional fields in terms of their zero modes and the KK excitations as follows:

\bea{sm_fer} l(x,y)  &  = & \frac{1}{\sqrt{\pi R}} \left\{ l^0_{L}(x) +
\sqrt{2} \sum_{n=1}^{\infty} \left[ l_L^n(x)  \cos \left(\frac{n
y}{R} \right)
 + l_R^n(x)  \sin \left(\frac{n y}{R} \right) \right] \right\},
\eea

\bea{single}
e_R(x,y)  &  = & \frac{1}{\sqrt{\pi R}} \left\{ e^0_{R}(x) + \sqrt{2}
\sum_{n=1}^{\infty} \left[ e_R^n(x)  \cos \left(\frac{n y}{R} \right)
 + e_L^n(x)  \sin \left(\frac{n y}{R} \right) \right] \right\},
\eea

\bea{higgs}
H(x,y)  & = & \frac{1}{\sqrt{\pi R}} \left\{ H^0 (x) + \sqrt{2} \sum_{n=1}^{\infty} H^n (x) \cos \left(\frac{n y}{R} \right) \right\}.
\eea

Note that the $l_L(x,y)$, $e_R(x,y)$ and $H(x,y)$ are even under $y \rightarrow -y$,
while $l_R(x,y)$ and $e_L(x,y)$ are odd under $y \rightarrow -y$.
The KK expansion for the SM singlet neutrino $N(x,y)$ is given by:
\bea{rh_neutr}
N(x,y)  &  = & \frac{1}{\sqrt{\pi r}} \left\{ N^0_{R}(x) + \sqrt{2}
\sum_{n=1}^{\infty} \left[ N_R^n(x)  \cos \left(\frac{n y}{r} \right)
 + N_L^n(x)  \sin \left(\frac{n y}{r} \right) \right] \right\}.
\eea
We have chosen only $N_R(x,y)$ to have zero mode, in analogy with that in the usual 4D where we introduce only right handed SM singlet to obtain a small neutrino mass. The quarks in the first family are also assumed to be confined on the fat brane of size $R$, with $R^{-1} \sim TeV$. The SM gauge bosons are also confined to this fat brane as in the UED, and will have their KK excitations in the $TeV$ scale satisfying the current experimental bounds.

Let us now discuss the charged lepton and neutrino masses in the effective four dimensional theory in our model. The charged lepton and Dirac neutrino mass terms arise from the Yukawa interactions:
\bea{inter}
S^{int}_1=\int d^{4}x \int^{\pi R}_0 dy [y_5\overline{l}(x,y)H(x,y)e(x,y) + \tilde{y}_5 \overline{l}(x,y)\tilde{H}(x,y)N(x,y)] + h.c.
\eea
where $\tilde{H}(x,y)= i \sigma_2 H^{*}(x,y)$, and $y_5$ and $\tilde{y}_5$ are five dimensional Yukawa couplings.

We assume that only the zero mode of the Higgs, $H^0(x)$ acquires vacuum expectation value which breaks the electroweak symmetry. This will then give rise to Dirac masses to the charged lepton as well as the light neutrino. From eq (\ref{inter}), using the KK expansions of eqs (\ref{sm_fer}-\ref{higgs}), we obtain:
\bea{inter2}
S^{int}_1=\int d^{4}x [\frac{1}{\sqrt{\pi R}}y_5\overline{e}_L e_R \frac{1}{\sqrt{2}}(v+h) + \frac{1}{\sqrt{\pi r}}\tilde{y}_5 \overline{\nu}_LN_R\frac{1}{\sqrt{2}}(v+h)] + h.c.
\eea
where $v \over \sqrt(2)$ is the vacuum expectation value of the Higgs field.

Defining the four dimensional Yukawa couplings $y_4$ and $\tilde{y}_4$ to be:
\bea{yuk_rel}
y_4 \equiv \frac{y_5}{\sqrt{\pi R}}, ~~\tilde{y}_4 \equiv \frac{\tilde{y}_5}{\sqrt{\pi r}}
\eea
we obtain:
\bea{mase}
m_e = y_4 \frac{v}{\sqrt{2}},~~ m_{\nu_e} = \tilde{y}_4 \frac{v}{\sqrt{2}}.
\eea

Note that even though the five dimensional Yukawa couplings, $y_5$ and $\tilde{y}_5$ are of the same order, the effective four dimensional neutrino Yukawa coupling is hierarchically small compared to that of the charged leptons because of the large size of the extra dimension, $r$ compared to the fat brane size, $R$. The SM singlet neutrino, N has a much larger space, $r$ available compared to $e_R$, thus producing tiny effective four dimensional Yukawa coupling, $\tilde{y}_4$ compare to $y_4$.

From eq. (\ref{yuk_rel}) and ({\ref{mase}}), we note that:
\bea{ratio}
\frac{m_{\nu_e}}{m_e}= \left(\frac{\tilde{y}_5}{y_5} \right)\sqrt{\frac{R}{r}}.
\eea

With $R^{-1} \sim 1~TeV$, and $r \sim submm \sim 10^{-3}~eV$, we obtain from eq. (\ref{ratio})(assuming $\tilde{y}_5$ and $y_5$ to be the same), $m_{\nu_e} \sim \sqrt{\frac{R}{r}}~m_e \sim 1.5 \cdot 10^{-2}~ eV.$
This is in the right range as needed for the solar neutrino experiment.

In the above calculation, we have included only the zero mode of $N(x,y)$, namely $N^0_R$. Since N propagates in the submm size dimension r, it will have KK excitations with mass of $10^{-3}~eV$ and its integral multiples. These will then mix substantially with the left-handed neutrino, affecting its SM coupling and thus contradicting experiments. To solve this problem, we introduce a bulk SM singlet scalar field $\Phi(x,y)$ which is odd under $y \rightarrow -y$. Then, the KK expansion for $\Phi(x,y)$ takes the form:
\bea{phiex}
\Phi(x,y)  & = & \sum_{n=1}^{\infty} \Phi^n (x) \sin \left(\frac{n y}{r} \right).
\eea

This gives rise to additional bulk Yukawa interactions:
\bea{bulk}
S^{1}_1= \tilde{Y}_{5} \int d^{4}x \int^{\pi r}_0 dy~\overline{N}(x,y)~N(x,y)~\Phi(x,y).
\eea
For simplicity, we assume that only $\Phi^{(1)}$ has $vev$ in the 5D Plank scale which we take to be $TeV$ or higher. Note that all our interactions conserve lepton number. No Majorana mass is introduced, and all masses are Dirac masses.

From eq. [10] and eq. (\ref{bulk}), we obtain the effective four dimensional action:
\bea{bulkred}
S^{1}_1=~M~\overline{N}^n_L~N^m_R~\left[\delta_{n-m-1,0} + \delta_{n-m+1,0} \right]
\eea
where $M \equiv \frac{\tilde{Y}_{5}<\Phi>}{\sqrt{\pi r}}$ and $<\Phi>$ is the vacuum expectation value of the singlet field $\Phi^{(1)}(x)$ in the $TeV$ scale or higher. We expect this mass, M, arising from the five dimensional SM singlet bulk fields $N$ and $\Phi$ to be in the string or five dimensional Plank scale. Including  this SM singlet Dirac mass, given by eq.(\ref{bulkred}) into account, we get a mass matrix involving the zero modes $\left(\overline{\nu}^0_{eL},~\overline{N}^1_L,~\overline{N}^2_L, ... \right)$ and $\left(N^0_R,~N^1_R,~N^2_R ... \right)$. Taking only the first two KK excitations of $N_L$ and $N_R$, the mass matrix is:
\bea{massmatr}
(\overline{\nu}^0_{eL},~\overline{N}^1_L,~\overline{N}^2_L)~
\left(\begin{array}{ccc}
m & \sqrt{2}m & \sqrt{2}m\\
0 & m' & M\\
0 & M  & 2m'
\end{array}\right)~
\left(\begin{array}{c}
N^0_R\\
N^1_R\\
N^2_R
\end{array}\right)\
\eea
where $m' \equiv \frac{1}{r}$.

The mass matrix in (\ref{massmatr}) is diagonalized by a bi-unitary transformation. The three mass eigenvalues are $\sim m,M,M$, and to order $m/M$, the corresponding eigenstates are :
\bea{eigenvect}
\left(\begin{array}{c}
1\\
O(m/M)\\
O(m/M)
\end{array}\right),~
\left(\begin{array}{c}
O(m/M)\\
1\\
1
\end{array}\right),~
\left(\begin{array}{c}
O(m/M)\\
-1\\
1
\end{array}\right)
\eea

Two Weyl neutrinos, $\nu^0_L$ and $N^0_R$ forms a 4-component massive Dirac neutrino of mass approximately m, where the $(N^1_L,~N^1_R)$ and $(N^2_L,~N^2_R)$ form two massive Dirac neutrinos $N^1$ and $N^2$ with very large $TeV$ scale mass $\sim M$. Note that while there is large mixing between $N^1_L$ and $N^1_R$, or  $N^2_L$ and $N^2_R$; there is only a tiny mixing of order m/M of $N^0_L$ with $N^1_L$ or $N^2_L$ and $N^0_R$ with $N^1_R$ or $N^2_R$. With $m \sim 10^{-2}~eV$, and $M \sim few~TeV$, this mixing is negligibly small to affect the $\nu_{eL}$ coupling in the SM model.

The same conclusion holds if we include the higher level KK excitations of N. If we include KK excitations of N up to the $n^{th}$ level, then the mass matrix involving $\left(\overline{\nu}^0_{eL},~\overline{N}^1_L,~\overline{N}^2_L, ..., ~\overline{N}^n_L \right)$ and $\left(N^0_R,~N^1_R,~N^2_R ...,~N^n_R \right)$ gives one light eigenvalue with mass $ \sim m $ and $(n-1)$ eigenvalues of mass $ \sim M $ in the $TeV$ scale or higher. Note that the Weyl neutrinos $\nu^0_{eL}$ and $ N^0_R $ form a light Dirac neutrino; $\nu_e$ of mass $ \sim m $. The KK excitations $N^i_L$ and $N^i_R$ form a massive Dirac neutrino $N^i$ of mass $ \sim M $.

Thus our model produces the large hierarchy between the light neutrino and charged lepton masses as well as the right magnitude for the left-handed light neutrino mass.Also, in one family, we have only one light Dirac neutrino, and a large number of very heavy ($\sim TeV$) SM singlet Dirac neutrinos. The mixing between the light neutrino and the heavy singlet Dirac neutrino are highly suppressed (of order $m_{\nu}/M$ with $M \sim TeV$).

\section{Models with two and three families}

Let us now consider two families of fermions. As in the previous section, we restrict ourself to only one extra dimension. The left handed doublets are $l_e(x,y)$ and $l_{\mu}(x,y)$. In addition to N(x,y), we introduce a second SM singlet neutrino in the bulk, $N'(x,y)$. Then, we will have bulk interactions for the $2^{nd}$ family similar to eqs. (\ref{inter}) and (\ref{bulk}) with the replacements
\bea{*}
l(x,y) \rightarrow l_{\mu}(x,y), e(x,y) \rightarrow \mu(x,y),
N(x,y) \rightarrow N'(x,y).\nonumber
\eea
 We also have the additional bulk interactions:
\bea{inter2pr}
S_{12}=\int d^{4}x \int^{\pi R}_0 dy[ y_{5(e \mu)} \overline{l}_e(x,y) H(x,y) \mu(x,y)
+ \tilde{y}_{5(e \mu)} \overline{l}_e(x,y) \tilde{H}(x,y) N'(x,y) \nonumber \\
+  ( e \leftrightarrow \mu)] + h.c.
\eea
and
\bea{inter2bulk}
S_{12}= \tilde{Y}'_{5} \int d^{4}x \int^{\pi r}_0 dy~M'_0~\left[\overline{N}(x,y)N'(x,y)+ \overline{N}'(x,y)N(x,y) \right] \Phi(x,y).
\eea

Using the KK decompositions for $l_e(x,y)$, $l_{\mu}(x,y)$, $H(x,y)$, $N(x,y)$ and $N'(x,y)$ and integrating over y in eqs. (\ref{inter}) and (\ref{bulk}), and in similar eqs for the $2^{nd}$ family, as well as eqs. [15],[16] (\ref{inter2}) and (\ref{inter2pr}), we obtain the mass terms in the effective four dimensional theory involving $\left( \overline{\nu}^0_{eL},\overline{N}^1_L,\overline{N}^2_L, ...; \overline{\nu}^0_{\mu L},\overline{N'}^1_L,\overline{N'}^2_L, ...\right)$ and $\left( N^0_R,N^1_R,N^2_R ...; N'^0_R,N'^1_R,N'^2_R ...\right)$. Again, keeping the terms only up to $N^2_L$, $N'^2_L$ and $N^2_R$, $N'^2_R$, we get a 6x6 mass matrix . Choosing the basis to be
$ \overline{L} = ( \overline{\nu}^0_{eL},\overline{\nu}^0_{\mu L},\overline{N}^1_L,\overline{N'}^1_L,\overline{N}^2_L,\overline{N'}^2_L )$,
$R = (N^0_R,~N'^0_R,~N^1_R,~N'^1_R,~N^2_R,~N'^2_R )$
, we get
\bea{intmass}
S_{mass}=\int d^{4}x~\overline{L}~\mathscr{M}~R
\eea
where
\bea{mass6}
\mathscr{M} =
\left(\begin{array}{cccccc}
m & q & \sqrt{2}m & \sqrt{2}q & \sqrt{2}m & \sqrt{2}q\\
q & p & \sqrt{2}q & \sqrt{2}p & \sqrt{2}q & \sqrt{2}p\\
0 & 0 & m' & 0 & M & M''\\
0 & 0 & 0 & p' & M'' & M'\\
0 & 0 & M & M'' & 2m' & 0\\
0 & 0 & M'' & M' & 0 & 2p'
\end{array}\right)
\eea

In (\ref{intmass}), M's are the $TeV$ scale or higher masses coming from eqs. (\ref{bulk}) and (\ref{inter2bulk}), p is the analogue of m from eq (\ref{massmatr}) for the muon sector, q is the cross-term between the $e$ and $\mu$ sector coming from eq(\ref{inter2}). For simplicity,we have also assumed $\tilde{y}_{5(e \mu)}=\tilde{y}_{5(\mu e)}$.

The mass matrix $\mathscr{M}$ can be diagonalized by a bi-unitary transformation. Two of the eigenvalues are small with mass of order m and p. These are the masses of the two light neutrinos, $m_{\nu_e}$ and $m_{\nu_{\mu}}$. The other four eigenvalues are of order M, giving very heavy neutrinos which are essentially SM singlets. As in the one family model, the mixing between these light and heavy neutrinos are extremely small, of order $m/M$. Thus, the coupling of these light neutrinos to the gauge bosons are essentially same as in the SM. Two very important features of the eq (\ref{mass6}) to note are the following. The off-diagonal element q in the first $(2 \times 2)$ sector of $\mathscr{M}$ is of the same order as m and p. So, the light masses $m_{\nu_e}$ and $m_{\nu_{\mu}}$ are of the same order and thus there is no hierarchy between these two masses, as is possibly the case experimentally. Furthermore, since m, p, q are all of the same order, we naturally get large mixing between $m_{\nu_e}$ and $m_{\nu_{\mu}}$, in agreement with the experimental observation. Enlarging the mass matrix (\ref{mass6}) to include up to n-th KK excitations, we get two light Dirac neutrinos of mass $\sim m$ and $2n$ very heavy Dirac neutrinos of masses $\sim M$. Although we have included only the first two KK modes of N and N', this pattern persists if we include the other higher modes.

We can easily extend the model to include three families. The left handed doublets are $l_e(e,y)$, $l_{\mu}(x,y)$ and $l_{\tau}(x,y)$. The SM singlets are $N(x,y)$, $N'(x,y)$ and $N''(x,y)$. Then, writing the bulk interactions analogous to (\ref{inter2pr}) and (\ref{inter2bulk}) to include all the above fields, we get the three family mass matrix. Again, keeping the terms only up to  $N^2_L,N^2_R,N'^2_L,N'^2_R,N''^2_L,N''^2_R$, we get a $(9 \times 9)$ mass matrix. Choosing the basis as $ \overline{L} = ( \overline{\nu}^0_{eL},\overline{\nu}^0_{\mu L},\overline{\nu}^0_{\tau L},\overline{N}^1_L,\overline{N'}^1_L,\overline{N''}^1_L,\overline{N}^2_L, \overline{N'}^2_L,\overline{N''}^2_L )$,
$R = (N^0_R,~N'^0_R,~N''^0_R,~N^1_R,~N'^1_R,~N''^1_R,~N^2_R,~N'^2_R,~N''^2_R)$, 9x9 mass matrix becomes:
\bea{*}
\mathscr{M} =
\left(\begin{array}{ccccccccc}
m_0 & m_{01} & m_{02} & \sqrt{2}m_0 & \sqrt{2}m_{01} & \sqrt{2}m_{02} & \sqrt{2}m_0 & \sqrt{2}m_{01} & \sqrt{2}m_{02} \\
m_{01} & m_1 & m_{12} & \sqrt{2}m_{01} & \sqrt{2}m_1 & \sqrt{2}m_{12} & \sqrt{2}m_{01} & \sqrt{2}m_1 & \sqrt{2}m_{12} \\
m_{02} & m_{12} & m_2 & \sqrt{2}m_{02} & \sqrt{2}m_{12} & \sqrt{2}m_2 & \sqrt{2}m_{02} & \sqrt{2}m_{12} & \sqrt{2}m_2 \\
0 & 0 & 0 & m'_0 & 0 & 0 & M_0 & M_{01} & M_{02}\\
0 & 0 & 0 & 0 & m'_1 & 0 & M_{01} & M_1 & M_{12}\\
0 & 0 & 0 & 0 & 0 & m'_2 & M_{02} & M_{12} & M_2\\
0 & 0 & 0 & M_0 & M_{01} & M_{02} & 2m'_0 & 0 & 0\\
0 & 0 & 0 & M_{01} & M_1 & M_{12} & 0 & 2m'_1 & 0\\
0 & 0 & 0 & M_{02} & M_{12} & M_2 & 0 & 0 & 2m'_2
\end{array}\right)\nonumber
\eea

The mass matrix, $\mathscr{M}$ can be diagonalized by a bi-unitary transformation. Three of the eigenvalues are small, of the order of $\sim m$, while the other six eigenvalues are of the order $M \sim TeV$ scale or higher. The parameters in the mass matrix can be suitable chosen to obtain values of the light neutrino masses and mixings in the observed range. For example, one such choice is $m_0=0.00157$, $m_1=0.025$, $m_2=0.029$, $m_{01}=0.015$, $m_{02}=0$, $m_{12}=0.028$, $m'_0=m'_1=m'_2=0.001$, $M_0=10^{14}$, $M_1=2\ast10^{15}$, $M_2=8\ast10^{15}$, $M_{01}=10^{15}$, $M_{02}=1.5\ast10^{15}$, $M_{12}=6\ast10^{15}$ (all masses in $eV$).

Although we have included only the first two KK modes of $N$, $N'$ and $N''$, this pattern of three light neutrino masses, and the rest very heavy, persists if we include any number of KK modes.

The model produces the hierarchy of masses between $m_e$ and $m_{\mu}$ naturally if we assume that the $2^{nd}$ family lives in a fat brane of much smaller size. For three families, charged lepton mass ratios are essentially given by (assuming the bulk Yukawa couplings are of the same order):
\bea{ratios3}
m_e~:~m_{\mu}~:~m_{\tau} = \sqrt{\frac{1}{R_1}}~:~\sqrt{\frac{1}{R_2}}~:~\sqrt{\frac{1}{R_3}}
\eea
where $R_1$,$R_2$ and $R_3$ are the sizes of the fat branes for the three families. Using the experimental values of the masses in (\ref{ratios3}), we get :
\bea{numbers}
R^{-1}_1~:~R^{-1}_2~:~R^{-1}_3 \sim 1~TeV~:~10^4~TeV~:~10^6~TeV
\eea

\section{Phenomenological Implications}

There are several interesting phenomenological implications (of our model) which can be tested in the upcoming neutrino experiments and high energy colliders. The light neutrinos in our model are Dirac particles. So neutrinoless double beta decay is not allowed in our model. This is a very distinctive feature of our model compared to the traditional see-saw mechanism. In the see-saw model, light neutrinos are Majorana particles, and thus neutrinoless double beta decay is allowed. Current limit on the double beta decay is $m_{ee} \sim 0.3 eV$. This limit is expected to go down to about $m_{ee} \sim 0.01 eV$ in future experiments \cite{betabeta}. If no neutrinoless double beta decay is observed to that limit, that will cast serious doubts on the see-saw model. In our model, of course, it is not allowed at any level.

Another interesting feature of our model is for the observation of the Kaluza-Klein (KK) excitations of the SM particles at the high energy colliders, such as LHC. In the ADD scenario, the SM particles are confined to a four dimensional wall ($D_3$ brane). So, no KK excitations of the SM particles exist. In the universal extra dimensions scenario \cite{acd}, all the SM particles propagate into the extra dimensions, and the current limit on the compactification scale from Tevatron collider is about 350 $GeV$ \cite{acd},\cite{matchev},\cite{mmn}. So, if the compactification scale is few $TeV$ or lower, we would observe the KK excitations of all the SM particles at the upcoming LHC. In contrast, in our model, the SM particles live in fat branes. The sizes of the fat branes are of different values for the three families. If the large extra dimension, r is of submm size, then to obtain the right value of the electron neutrino mass, eq. (\ref{ratio}), we get the size of the fat brane for the first family, $R^{-1}_1 \sim 1~ TeV$, while the other sizes for the $2^{nd}$ and $3^{rd}$ families are around $10^4~TeV$ and $10^6~TeV$ respectively. Thus, unlike in the UED model, only the KK excitations of the first family will be observed at the LHC, but not those of the $2^{nd}$ and $3^{rd}$ families. Thus is a very distinguishing feature of our model compared to UED and ADD models.

In models in which the SM particles propagate into the extra dimensions, the unification of the three gauge couplings are accelerated due to the contribution of the KK modes of the SM particles \cite{power}. Above the compactification scale, the evolution of the gauge couplings become power law, instead of logarithmic as in the 4D case. In our model, since KK excitation of the $1^{st}$ family has much lower compactification scale (order $TeV$) compared to the second and third family, the unification will take place at a higher scale than in UED.

One interesting question to ask in our model is where do the SM gage bosons live?
Since the three SM families live in fat branes of three different sizes, are their gauge coupling universal (i.e. same for all three families)? To be specific, let us assume that the SM gauge bosons live in a fat brane of size $R'$, where $R' \geq R_i$. Then, their coupling to the SM fermions of the $i^{th}$ family is given by:
\bea{boso}
S^i_{gauge} & = & \int d^{4}x \int^{\pi R_i}_0 dy \frac{1}{(\sqrt{\pi R_i})^2 \sqrt{\pi R'}} g_5 \overline{f_i}(x,y) \Gamma^{\mu} T^a f_i(x,y) A^a_{\mu} \nonumber\\
            & = & \int d^4x \frac{g_5}{\sqrt{\pi R'}}[\overline{f_i}^0 \gamma^{\mu} T^a f^0_i A^{a0}_{\mu} + KK ~terms]
\eea

Note that as long as $R_i$ is smaller than $R'$, the four dimensional gauge couplings are universal for all three families. One very interesting implication of this is that the compactification scale for the SM gauge bosons are smaller than the sizes of the fat branes for all three SM fermionic families. Thus, in our model, the KK excitations of the gauge bosons are lighter than the KK excitations of the fermions, and will be the first ones to be observed at the LHC.

There are several other features of our model that can be tested in the high energy colliders such as the Tevatron or LHC. As in the UED, KK particles can only be pair produced in our model because of the KK number conservation. however, only the KK excitations of the gauge bosons and the first family of fermions are accessible. The KK excitations of the $2^{nd}$ and $3^{rd}$ families are not accessible  even at LHC. Thus, the number of final states in the production processes is significantly reduced. This effect is even more because of the doubling of the KK states. Because of this reduction in the production cross section, current Tevatron bound of $350 GeV$ \cite{acd}, \cite{matchev},\cite{mmn} on the compactification scale will be reduced further. Also, in our scenario, KK excitations of the gluons are lighter than the KK excitations of the quarks. One loop correction to the KK masses will be somewhat larger for the KK gluons than those of the KK quarks. However, if the $R'^{-1}$ is significantly smaller than $R^{-1}_1$, this correction may not be enough (this is in contrast to UED where g* is always heavier than q*). This will have interesting implications for their decays and the final state collider signals.

\section{Conclusions}

We have proposed a model for the charged leptons and fermion masses in the context of extra dimensions. The three families of fermions live in fat branes of three different sizes of $TeV^{-1}$ or smaller, while the gravity and the SM singlet neutrinos propagate in the full extra dimension of submm size. This is different from the original ADD model where all the SM particles are confined to a wall of zero thickness. The hierarchy of the charged lepton masses is accommodated using the hierarchy of the size of the fat branes. The neutrino masses of all the three families come from the submm size, and hence not hierarchical. The neutrino mass of a family is related to the corresponding charged lepton mass by the square root of the fat brane size to the submm size. This naturally explains why the neutrino masses are so much smaller compared to the charged leptons. When extended to more than one family, large mixing in the neutrino sector is naturally expected in this model. One key distinguishing feature compared to the see-saw model is that here light massive neutrinos are Dirac particles, and thus no neutrinoless double beta decay are expected at any level. This clearly distinguishes our model from the see-saw, and can be tested in the future high precision neutrinoless double beta decay experiments. The model has several interesting collider signatures that can be tested at LHC, such as only the KK excitations of the first family in within LHC reach, and also KK excitations of the gauge bosons will be lighter than those of the fermions.

\section*{Acknowledgments}

We thank K.S. Babu for useful discussions. This work was supported in part by the US Department of Energy, Grant Numbers DE-FG02-04ER41306 and DE-FG02-04ER46140.

\end{document}